\documentclass[12pt,a4paper]{iopart}

\usepackage{iopams}							
\usepackage{setstack}						
\usepackage[english]{babel}			
\usepackage{graphicx}						
\usepackage{acronym}						
\usepackage{hyperref}						
\usepackage{cite}								
\usepackage{color}

\newcommand{\ii}{\rmi}										
\newcommand{\ee}{\rme}										
\renewcommand{\Re}{{\rm Re}\,}						
\renewcommand{\Im}{{\rm Im}\,}						
\newcommand{\PT}{$\mathcal{PT}$}
\newcommand{\pmsa}{{\color{red}\pm}}

\newcommand{\pmsb}{{\color{green}\pm}}
\newcommand{\mpsb}{{\color{green}\mp}}

\begin{document}

\acrodef{BEC}{Bose-Einstein condensate}
\acrodef{GPE}{Gross-Pitaevskii equation}
\acrodef{TDVP}{time-dependent variational principle}

\title{Realizing \texorpdfstring{\PT}{PT}-symmetric BEC subsystems
	in closed	hermitian systems}
\date{}
\author{Robin Gut\"ohrlein, Jan Schnabel, Ibrokhim Iskandarov, Holger
	Cartarius, J\"org Main and G\"unter Wunner}
\address{1. Institut f\"ur Theoretische Physik,
  Universit\"at Stuttgart, 70550 Stuttgart, Germany}
\ead{Robin.Gutoehrlein@itp1.uni-stuttgart.de}

\begin{abstract}
	In open double-well Bose-Einstein condensate systems which balance in- and
	outfluxes of atoms and which are effectively described by a non-hermitian
	\PT-symmetric Hamiltonian \PT-symmetric states have been shown to exist.
	\PT-symmetric states obey parity and time reversal symmetry.  We tackle the
	question of how the in- and outfluxes can be realized and introduce a
	hermitian system in which two \PT-symmetric subsystems are embedded. This
	system no longer requires an in- and outcoupling to and from the environment.
	We show that the subsystems still have \PT-symmetric states. In addition we
	examine what degree of detail is necessary to correctly model the
	\PT-symmetric properties and the bifurcation structure of such a system.  We
	examine a four-mode matrix model and a system described by the full
	Gross-Pitaevskii equation in one dimension.  We see that a simple matrix model
	correctly describes the qualitative properties of the system. For sufficiently
	isolated wells there is also quantitative agreement with the more advanced
	system descriptions.	We also investigate which properties the wave functions
	of a system must fulfil to allow for \PT-symmetric states. In particular the
	requirements for the phase difference between different parts of the system
	are examined.
\end{abstract}


\section{Introduction}
\label{sec:introduction}

In conventional quantum mechanics hermitian operators are used to describe
closed quantum systems. These operators allow only for real eigenvalues, which
can represent physical observables. Since systems in the real world are hardly
ever completely isolated, the environment must be taken into account. Due to a
lack of knowledge about the actual layout of the environment of a system or
because the environment is too complicated to be completely calculated, one can
effectively describe such systems as open quantum systems as long as the
interaction to the environment is known. Such Hamiltonians are often no longer
hermitian.  The interaction with the environment, e.g.\ gain and loss of the
probability amplitude, that is the wave function, can be expressed by complex
potentials \cite{Moiseyev2011a}. These Hamiltonians in general do not have a
real eigenvalue spectrum.

A special class of non-hermitian operators was investigated by Bender and
Boettcher in 1998 \cite{Bender98}. For certain parameter ranges these operators
also had purely real eigenvalue spectra. The origin of the special property can
be traced back to the \PT-symmetry of the operator, where the \PT-operator
consists of the parity operator $\mathcal{P}$ and the time reversal operator
$\mathcal{T}$. The parity operator exchanges $\hat x \rightarrow -\hat x$ and
$\hat p \rightarrow -\hat p$. The time reversal operator replaces $\hat p
\rightarrow -\hat p$ and $\ii \rightarrow -\ii$. A \PT-symmetric system has a
Hamiltonian which fulfils $[H, \mathcal{PT}] = 0$.  For a system with
\begin{equation}
	H =	-\Delta + V
\end{equation}
the position space representation of the potential must obey the condition
\begin{equation}
	V(x) = V^*(-x), \label{eq:ptsymmetrycondition}
\end{equation}
i.e.\ the real part of the potential must be an even function in the spatial
coordinate and the imaginary part must be an odd function.  \PT-symmetric
systems have been studied theoretically for quantum systems
\cite{qm1,qm2,Mehri,Bender1999a}. However, the concept of \PT-symmetry is not
restricted to quantum mechanics. Indeed, the experimental breakthrough was
achieved in optical wave guides by R\"uter \etal \cite{Rueter10} when in such a
system the effects of \PT-symmetry and \PT-symmetry breaking were observed. This
has led to a still increasing interest in the topic \cite{PhysRevA.88.053817,
Deffner2015, Albeverio2015, Mostafazadeh2013b}, and \PT-symmetric systems have
also been studied in microwave cavities \cite{Bittner2012a}, electronic devices
\cite{Schindler2011a,Schindler2012a}, and in optical \cite{0305-4470-38-9-L03,
Guo09, ramezani10, musslimani08a, optic1, optic2, Makris2010a, makris08,
Chong2011, Peng2014} systems.  Also in quantum mechanics the stationary
Schr\"odinger equation was solved for scattering solutions \cite{qm2} and bound
states \cite{Mehri}. Note that it was shown in \cite{Dast2014} that the
characteristic \PT-symmetric properties are still found when a many-particle
description is used.

In \cite{Klaiman08} it was suggested that \PT-symmetry could also be realized
in quantum systems, namely in Bose-Einstein condensates (BECs).  The BEC would
be captured in a symmetric double-well potential where particles are gained in
one well and lost in the other. This loss and gain can then be described
by a complex potential coupling the system to the environment.

The time-independent solutions (see \ref{s:tigpe}) of such a \PT-symmetric
double-well system can in the simplest possible case \cite{Graefe12} be
described by the matrix
\begin{eqnarray}
	\left( \begin{array}{cc}
	    -g | \psi_1 |^2 - \ii \gamma & v \\
			v & -g | \psi_2 |^2 + \ii \gamma
	  \end{array} \right) 
   \left( \begin{array}{c} \psi_1 \\ \psi_2 \end{array} \right)
		=
  	\mu \left( \begin{array}{c} \psi_1 \\	\psi_2 \end{array} \right),
	\label{eq:matrix2}
\end{eqnarray}
where $\psi_1$ and $\psi_2$ represent the occupations of the two wells with
atoms in the condensed phase and $\mu$ is the chemical potential.  This
description can be derived from a non-hermitian representation of a
many-particle Bose-Hubbard dimer \cite{Graefe08a}. The off-diagonal elements $v$
of the matrix describe the couplings between the wave functions in the two
potential wells. The diagonal contains a nonlinear entry introducing the
particle-particle interaction described by an s-wave scattering process. Its
strength can be changed via the parameter $g$ which is proportional to the
s-wave scattering length and its physical variation can be achieved close to
Feshbach resonances. In comparison to the original model from \cite{Graefe12}
the replacement $g \rightarrow - g$ is introduced to be consistent with the
other models which will be shown later on.  In addition the diagonal contains an
imaginary term with the parameter $\gamma$. This term models a particle gain in
one well and a particle loss in the other.  This gain and loss is provided by
the (not further described) environment. The wave functions consist of two
complex values and contain no spatial information.  Therefore the parity
operator $\mathcal{P}$, which normally exchanges $\hat x$ with $-\hat x$,
exchanges $\psi_1$ with $\psi_2$ and vice versa. It is also assumed that the
potential wells are isolated enough such that the nonlinear interaction between
$\psi_1$ and $\psi_2$ can be neglected.

\begin{figure}
  \noindent\includegraphics[width=0.99\textwidth]{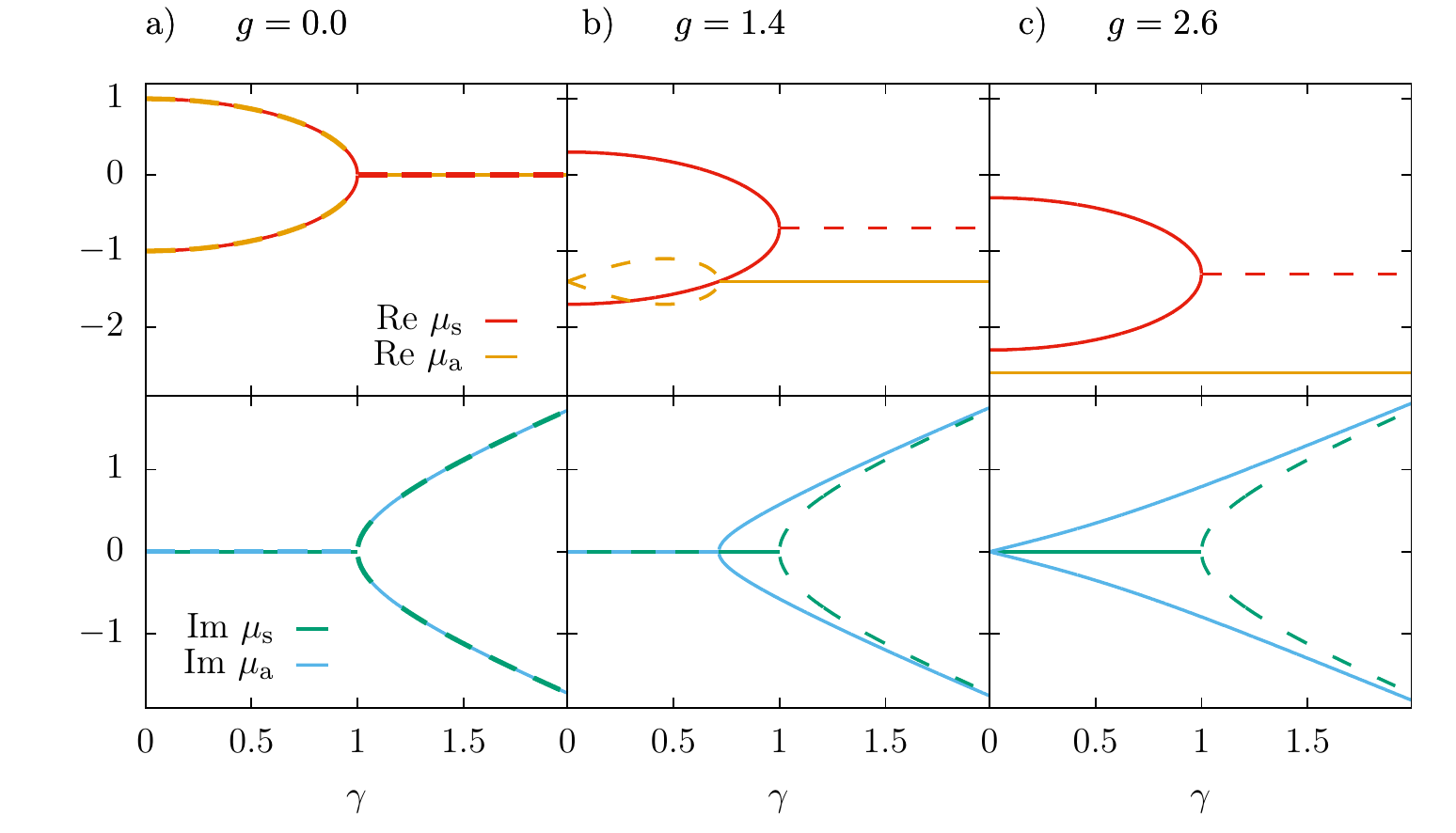}
	\caption{ Analytic solutions for the chemical potential \eref{eq:matrix2ana}
		of the two-dimensional matrix model described in \eref{eq:matrix2}. The
		coupling strength $v=1$, and	the nonlinearities $g = 0$ in a), $g = 1.4$ in
		b) and $g = 2.6$ in c) are used. The analytically continued solutions are
		plotted using dashed lines.}
	\label{fig:2dmatrix}
\end{figure}

The system \eref{eq:matrix2} is solved analytically \cite{Graefe12} for wave
function vectors $\psi$ which are normalized to one. The chemical potential
reads
\begin{eqnarray}
  \mu_s &= -\frac{g}{2} \pm \sqrt{v^2 - \gamma^2}, \nonumber \\
	\mu_a &= -g \pm \gamma \sqrt{\frac{4v^2}{g^2+4\gamma^2}-1}.
	\label{eq:matrix2ana}
\end{eqnarray}
The values $\mu_s$ in \eref{eq:matrix2ana} are the \PT-symmetric solutions, and
the \PT-broken solutions of the system are denoted $\mu_a$. All solutions
are shown in \fref{fig:2dmatrix}. For small $\gamma$ the system without
nonlinearity ($g = 0$) shows only \PT-symmetric states with real chemical
potential $\mu \in \mathbb{R}$ as can be observed in \fref{fig:2dmatrix}a.
These states pass through a tangent bifurcation at $\gamma = \gamma_c = 1$, and
two \PT-broken states emerge. For $\gamma > \gamma_c$ only \PT-broken states
with a complex chemical potential $\mu \in \mathbb{C}$ exist. 

For a nonlinearity $g > 0$ the bifurcation in which the two \PT-broken states
are created moves to a smaller value of $\gamma$ on one of the \PT-symmetric
branches (compare \fref{fig:2dmatrix}b). A pitchfork bifurcation is formed.
Thus for nonzero values of $g$ there is an additional parameter region for
$\gamma$, in which \PT-symmetric and \PT-broken states exist simultaneously.
When the nonlinearity is increased even further ($g > 2$) we see in
\fref{fig:2dmatrix}c that the pitchfork bifurcation is no longer present and the
\PT-broken states exist for all values of $\gamma$. A thorough examination of
the bifurcation structure and of the associated exceptional points can be found
in \cite{Heiss13a}.

The matrix model does not take the spatial extension of the system into account.
In general BECs can be described by the nonlinear Gross-Pitaevskii equation
\cite{Pitaevskii03a}. Often $\delta$ functions have been used to gain a deeper
insight \cite{Mostafazadeh2006a,Jones2008a,Mostafazadeh2009a,qm2,Mehri,
Mayteevarunyoo2008a,Rapedius08a,Wit08a,Fassari2012a,Demiralp2005a,
Jones1999a,Ahmed2001a,Uncu2008a}. Therefore a simple model to include spatial
effects describes the potential with double-$\delta$ functions
\cite{Cartarius12b}. In this system two $\delta$-wells exist at the positions $x
= \pm b$. While both of these wells have the same real depth they possess
antisymmetric imaginary parts. That is, one well has a particle gain and the
other has an equally strong particle drain. The potential fulfils the
\PT-symmetry condition \eref{eq:ptsymmetrycondition} and the corresponding
Gross-Pitaevskii equation is
\begin{eqnarray}
 \fl	-\psi''(x) -\left[ (1+\ii\gamma)\delta(x+b)
		+(1-\ii\gamma)\delta(x-b) \right] \psi(x)- g | \psi(x)|^2 \psi(x)
		= \mu \psi(x) \label{eq:delta2}.
\end{eqnarray}
In this system \PT-symmetric solutions and \PT-symmetry breaking were found.

In \cite{Dast13, Haag14} a similar two well system was examined in much greater
detail by using a more realistic potential well shape. The Gross-Pitaevskii
equation of such a BEC can be written as
\begin{eqnarray}
 (-\Delta + V(x) - g|\psi(x,t)|^2) \psi = \mu \psi \label{eq:gauss2} 
\end{eqnarray}
with the complex potential
\begin{eqnarray}
	V(x) = \frac{1}{4} x^2 + V_0^{\rm G} \ee^{-\sigma x^2} + \ii \gamma x \ee^{-\rho x^2}
	\quad{\rm with}~
	\rho = \frac{\sigma}{2 \ln(4 V_0^{\rm G} \sigma)}
	\label{eq:gauss2_pot} 
\end{eqnarray}
containing the BEC in a harmonic trap divided by a Gaussian potential barrier
into two wells. The parameter $\rho$ is chosen in such a way that the maximal
coupling between the subsystems occurs at the minima of the potential wells.
The stationary states show the same general behaviour as those in the matrix
model.

All descriptions so far used complex potentials to effectively describe the
environment. Therefore only the \PT-symmetric part of the whole system was
described in detail while the concrete layout of the environment itself was not
specified. We will now discuss how it might be possible to embed such a
\PT-symmetric two-well system into a larger hermitian system and therefore
explicitly include the environment into our description.

As a first step in this direction a hermitian four well model was used
\cite{Kreibich2013a, Kreibich2014}, where the double-well with in- and outgoing
particle fluxes is achieved by embedding it into the larger system. The two
outer wells have time-dependent adjustable parameters namely the potential
depth and the coupling strength to the inner wells. By lowering and raising
these wells a particle gain and loss in the two inner wells can be obtained,
which exactly corresponds to the loss and gain in the non-hermitian two-well
model.  However, the \PT-symmetric subsystem of the inner wells loses its
properties when the well which provides the particle gain is depleted.  A
second possible realization was suggested in \cite{Single2014}, where the wave
function of a double-well potential was coupled to additional unbound wave
functions (e.g.  one ingoing and one outgoing) connecting the gain and loss of
the system with a reservoir. These auxiliary wave functions replace the
previously unknown environment of the system.

In this paper we propose an additional way of realizing a \PT-symmetric
two-well system by extending the approach used in \cite{Single2014}. We couple
two \emph{stationary} bound wave functions, where each of them has the shape of
that of the corresponding \PT-symmetric system and their combination results
in a hermitian system. The influx from one system originates from the second
and vice versa. By tuning the coupling strength between the two systems we will
be able to control the gain and loss in the subsystems. In contrast to
\cite{Single2014} our systems are closed and do not require incoming or
outgoing wave functions or time-dependent potentials. We will show that for
suitable states the subsystems are indeed \PT-symmetric, however, also
\PT-symmetry breaking can be observed. 

In \sref{sec:2} a four-dimensional matrix model will be constructed similar to
the model \eref{eq:matrix2} to observe the general structure of the eigenstates
and to determine their \PT-symmetric properties. For this model analytical
solutions can be found. In a next step a Hamiltonian is constructed to combine
two subsystems with a spatial resolution in one dimension for the wave function
similar to the double-$\delta$-potential used in \eref{eq:delta2}. In these
systems effects which depend on the shape of the wave functions can be
observed. We will examine which detail of description is necessary to capture
the \PT-symmetric properties of the system and the bifurcation structure. Since
a model with double-$\delta$-potentials is only a rough approximation of the
reality we will also introduce an additional system. This system is constructed
by coupling two subsystems of the form \eref{eq:gauss2}. It not only has an
expanded wave function, which resolves spatial information, but also possesses
more realistic extended potential wells. In addition the coupling between the
two modes takes place over an extended area of space and is not confined to the
locations of the $\delta$-wells.

Subsequently we will compare the results obtained with the different
descriptions in \sref{sec:3}. We will also compare the bifurcation structure
with the model \eref{eq:matrix2}. In addition the influence of the phase
difference between the two subsystems on the stationary states will be
determined. A summary and discussion of the results is given in \sref{sec:4}.

\section{Coupling of two two-well potentials in one hermitian system}
\label{sec:2}

\begin{figure}
  \noindent\includegraphics[width=0.99\textwidth]{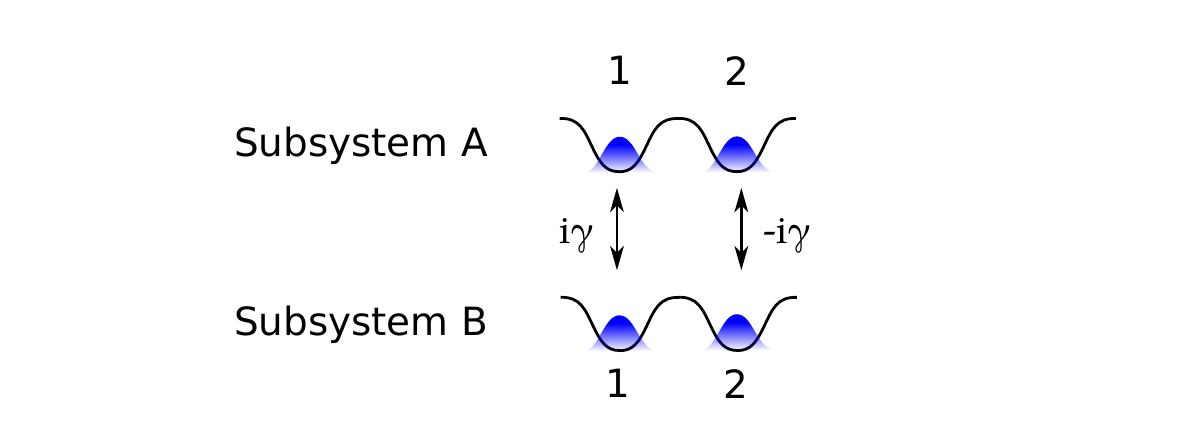}
	\caption{ This sketch illustrates how two double-well subsystems are combined
		into a closed hermitian system. The coupling and description of the wells
		is given with a varying degree of detail for the different systems
		discussed in this paper.}
	\label{fig:model}
\end{figure}

In \fref{fig:model} the layout of two coupled two-well systems is sketched.
The two subsystems are labelled A and B and each contains two wells with the
labels 1 and 2. In the drawing the potentials of the wells are extended. This
corresponds to an ansatz as shown in \eref{eq:gauss2} and \eref{eq:gauss2_pot}
and will be one of the systems studied in this work. Each of the wells is
coupled to its counterpart in the other subsystem. The coupling strength is
described by the parameter $\gamma$. Since the strength of the in- and
outcoupling is also determined by the wave function of the other subsystem,
\PT-symmetry can only exist for both subsystems.  There is no \PT-symmetry for
arbitrary states but only for states with an appropriate symmetry.  As mentioned
in the introduction we will investigate the setup in three different degrees of
detail.

There exist various other systems which have four distinguished modes. A family
of such systems named plaquettes was examined \cite{1751-8121-45-44-444021,
Yang:14}.  These systems are seen as a first step towards building \PT-symmetric
lattice systems \cite{PhysRevA.85.033825, abdullaev11}. The plaquettes exist in
various configurations which differ in the coupling between the sites.  In
contrast to the model proposed in this paper the gain and loss in these
plaquettes is still provided by non-hermitian terms.

\subsection{Four-dimensional matrix model}

In a first step we construct the four-dimensional hermitian matrix model.
Therefore we place two matrices of the shape \eref{eq:matrix2} on the main
diagonal blocks in our new matrix $M$ and remove the terms which couple the
system to the environment. They are replaced with coupling terms in the 
off-diagonal $2\times2$ matrix-blocks, i.e.
\begin{eqnarray}
 M = \left( \begin{array}{cc|cc}
    -g | \psi_{{\rm A},1} |^2 & v & - \ii \gamma & 0 \\
    v & -g | \psi_{{\rm A},2} |^2 & 0 & + \ii \gamma \\
		\hline
    +\ii \gamma & 0 & -g | \psi_{{\rm B},1} |^2 & v \\
    0 & -\ii \gamma & v & -g | \psi_{{\rm B},2} |^2
  \end{array} \right) \label{eq:matrix4}
\end{eqnarray}
with the wave function
\begin{equation}
	\psi = \left( \psi_{{\rm A},1} , \psi_{{\rm A},2} ,
								\psi_{{\rm B},1} , \psi_{{\rm B},2} \right).
\end{equation}
The elements of $\psi$ are four complex values with no information about the
spatial extension of the wave function. The first two values $\psi_{A,1},
\psi_{A,2} \in \mathbb{C}$ represent the wave function amplitudes in the 
double-well potential of subsystem A while the values $\psi_{B,1}, \psi_{B,2} \in
\mathbb{C}$ represent the amplitudes in the subsystem B. Therefore in this
context the parity operator $\mathcal{P}$ exchanges $\psi_{{\rm A},1}$ with
$\psi_{{\rm A},2}$ and $\psi_{{\rm B},1}$ with $\psi_{{\rm B},2}$.  The two
diagonal submatrices will form our subsystems A and B each with two wells
indicated by the indices 1 and 2. The first well of subsystem A is coupled via
$M_{1,3} = -\ii\gamma$ to the first well in subsystem B. The first well of
subsystem B is coupled via $M_{3,1} = \ii\gamma$ to $\psi_{{\rm A},1}$,
therefore keeping the matrix hermitian.  The second wells are coupled in a
similar manner but with opposite signs.  Note that the coupling terms do not
yet guarantee a symmetric gain and loss in a subsystem since the gain and loss
depend also on the value of the wave function of the other mode. 

The coupling between the potential wells in one subsystem is done via the
parameter $v$. The parameter $g$ still describes the particle-particle
scattering in one well, but no scattering between the overlap of the wave
functions from different wells is taken into account.

The time-independent equation describing the stationary states of the complete
system reads
\begin{eqnarray}
  M \psi = \mu \psi
  \label{eq:matrix} 
\end{eqnarray}
with real eigenvalues $\mu \in \mathbb{R}$ because the matrix $M$ is hermitian.
Since we are also interested in the \PT-symmetric properties of the subsystems
we extend \eref{eq:matrix} to
\begin{eqnarray}
  M \psi = \left( \begin{array}{cc}
		M_{\rm A} & M_{\rm C} \\
		M_{\rm C}^* & M_{\rm B}
  \end{array} \right)
  \left( \begin{array}{c}
 		\psi_{{\rm A}} \\
 		\psi_{{\rm B}} \\
	\end{array} \right)
	=
  \left( \begin{array}{c}
 		\mu_A \psi_{{\rm A}} \\
 		\mu_B \psi_{{\rm B}} \\
	\end{array} \right)
	\label{eq:complexmu}
\end{eqnarray}
with independent eigenvalues $\mu_i \in \mathbb{C}$ for both subsystems and 
\begin{eqnarray}
   M_{\rm C} = \left(\begin{array}{cc}
			-\ii \gamma & 0 \\
			0 & \ii \gamma
		\end{array}\right) ~, ~~
	M_i = \left( \begin{array}{cc}
	   -g | \psi_{i,1} |^2 & v \\
	    v & -g | \psi_{i,2} |^2 
		\end{array}\right)
\end{eqnarray}
for $i = \rm A, B$. We can interpret \eref{eq:complexmu} as two separate
equations for both subsystems where the gain and loss is provided by the other
subsystem via the matrix $M_{\rm C}$.  For $\mu_{\rm A, B} \in \mathbb{C}$ this
also allows for \PT-broken states where the norm of the subsystems is no longer
maintained, but is increased or decreased. Such solutions are therefore
non-stationary states, but because the particle number of the whole system is
conserved, there is no unlimited exponential growth or decay possible.
Therefore these solutions describe only the onset of their growing or decaying
temporal evolution. Only states with $\mu_{\rm A} = \mu_{\rm B} \in \mathbb{R}$
are stationary \PT-symmetric solutions. For $\mu_{\rm A} = \mu_{\rm B}^*$
\eref{eq:complexmu} leads to solutions where the gain and loss of subsystem A
(represented by $\Im \mu_A$) is compensated by the loss and gain of subsystem B
($\Im \mu_B$). Therefore the total particle number is indeed conserved.

We can parametrize the ansatz of the wave function for this model and reduce the
parameter count by removing a global phase. Solutions exist for different ratios
of the probability amplitude between the two subsystems, but they may not exist
over the whole range of the parameters. To simplify the equations we choose to
restrict the norm of each subsystem to one. This leads to the ansatz
\begin{eqnarray}
  \psi = 
  	\left( \begin{array}{c}
 			\psi_{{\rm A},1} \\
	 		\psi_{{\rm A},2} \\
	 		\psi_{{\rm B},1} \\
	 		\psi_{{\rm B},2} \\
	\end{array}	\right)
	=
	\left( \begin{array}{l}
	  \cos\theta_{\rm A} e^{+\ii \varphi_{\rm A}} \\
  	\sin\theta_{\rm A} e^{-\ii \varphi_{\rm A}} \\
	  \cos\theta_{\rm B} e^{+\ii \varphi_{\rm B} + \ii \varphi_{\rm rel}} \\
	  \sin\theta_{\rm B} e^{-\ii \varphi_{\rm B} + \ii \varphi_{\rm rel}}
	\end{array}	\right)
	\label{eq:matrixansatz}
\end{eqnarray}
with the parameters $\theta_A$ and $\theta_B$ determining the distribution of
the probability amplitude of the wave function on the two potential wells in
one subsystem and the parameters $\varphi_A$ and $\varphi_B$ describing the
phase difference. The parameter $\varphi_{\rm rel}$ defines the phase
difference between the two subsystems.  By applying additional symmetry
restrictions and thereby reducing the parameter count even further, analytical
solutions of \eref{eq:complexmu} can be obtained and are presented in
\sref{sec:3}. All other solutions can be gained numerically by applying a
multidimensional root search.

\subsection{Model with a spatial resolution of the wave function}

We want to know if the basic description provided by the matrix model is
sufficient to capture the \PT-symmetric properties and the bifurcation
structure of the system or if a more detailed description is necessary. We do
this in two steps. First we allow for the more detailed information of a
spatially resolved wave function but retain the concept of isolated coupling
points. The double-$\delta$ system keeps the mathematical and numeric intricacy
at bay but still provides a spatial resolution for the wave function. Therefore
we combine two systems with $\delta$-potentials \eref{eq:delta2} which describe
each subsystem in one spatial dimension. The subsystems are then coupled at the
positions of the $\delta$-wells $x = \pm b$.  The depth of the potentials is
controlled by $V_0^{\rm D}$. Both the depth $V_0^{\rm D}$ and the distance $2b$
between the wells correspond to the coupling parameter $v$ in the matrix
model. The coupling strength between the two subsystems is controlled by
$\gamma$ and is only present at the two points $x = \pm b$, i.e.\ the potential
has no spatial extension. The dimensionless coupled Gross-Pitaevskii equations
read
\begin{eqnarray}
  \fl \left[ -\frac{\partial^2}{\partial x^2}
       -g |\psi_{\rm A}|^2
			 + V_0^{\rm D} (\delta(x-b) + \delta(x+b)) \right] \psi_A \nonumber \\
			 + \ii \gamma \left[ \delta(x-b)\psi_B(b) - \delta(x+b)\psi_B(-b) \right]
	= \mu_A \psi_A, \nonumber\\
  \fl  \left[ -\frac{\partial^2}{\partial x^2}
       -g |\psi_{\rm B}|^2
			 + V_0^{\rm D} (\delta(x-b) + \delta(x+b)) \right] \psi_B \nonumber \\
			 - \ii \gamma \left[ \delta(x-b)\psi_A(b) - \delta(x+b)\psi_A(-b) \right]
	= \mu_B \psi_B, \label{eq:deltaGPEb}
\end{eqnarray}
with the same physical interpretation of $\mu_A$ and $\mu_B$ as in
\eref{eq:complexmu} for the matrix model.  Stationary states of the system are
calculated numerically exact by integrating the wave functions outward from $x
= 0$ and by imposing the appropriate boundary conditions on the wave functions.
We require that the wave functions have to approach zero when $x \rightarrow
\pm \infty$. For numerical purposes it is sufficient for the wave functions to
have small values at $x = \pm x_{\rm max}$,
\begin{eqnarray}
	\psi_{\rm A, B}(\pm x_{\rm max}) \approx 0. \label{eq:deltainf}
\end{eqnarray}
An additional condition can be required for the norm of the wave function. In
agreement with the normalized ansatz \eref{eq:matrixansatz} in the matrix model
we search for solutions that fulfill
\begin{eqnarray}
	||\psi_{\rm A, B}||^2 = 1. \label{eq:deltanorm}
\end{eqnarray}
Both wave functions are real at $x = 0$. With this we enforce a global phase
and the phase difference between the two modes at $x = 0$ to be $\varphi_{\rm
rel} = 0$.

The 10 (real) free parameters are $\Re \mu_{A,B}$, $\Im \mu_{A,B}$,
$\Re\psi_{\rm A,B}(0)$, $\Re \psi'_{\rm A,B}(0)$ and $\Im \psi'_{\rm A,B}(0)$.
They are chosen such that the 10 (real) conditions, i.e.\ the norm
\eref{eq:deltanorm} and the boundary conditions at $x = \pm x_{\rm max}$
\eref{eq:deltainf} are fulfilled. Note that there are no constraints on the
$\mu_{A,B}$.  We will see that for stationary \PT-symmetric solutions the
result is $\mu_A = \mu_B \in \mathbb{R}$. This is not a constraint on the root
search.

\subsection{Model with a spatial resolution of both the potential well and the
	coupling}

We consider an additional system and remove a further restriction, viz.\ the
point-like coupling approach, by duplicating the system from \eref{eq:gauss2},
where the wells are formed by a harmonic trap and divided by a Gaussian
potential barrier. This does not only provide us with a system with much more
realistic potential wells but also allows us to extend the coupling of the two
subsystems over the whole space. The time-independent GPEs of the system read
\begin{eqnarray}
	\Big( -\frac{\partial^2}{\partial x^2}
	- g | \psi_{\rm A} |^2
	+ \frac{1}{4}	x^2 + V_0^{\rm G} \ee^{-\sigma x^2} &\Big) \psi_{\rm A}
	+ \ii \gamma x \ee^{-\rho x^2} \psi_{\rm B}
	&= \mu_A \psi_{\rm A},
	\nonumber\\
		\Big( -\frac{\partial^2}{\partial x^2}
	- \underbrace{g | \psi_{\rm B} |^2}_{\rm contact}
	+ \underbrace{\frac{1}{4} x^2 + V_0^{\rm G} \ee^{-\sigma x^2}}_{\rm trap}
			&\Big)  \psi_{\rm B}
	- \underbrace{\ii \gamma x \ee^{-\rho x^2} \psi_{\rm A}}_{\rm coupling}
	&= \mu_B \psi_{\rm B}. \label{eq:gauss}
\end{eqnarray}
The parameter $V_0^{\rm G}$ controls the height of the potential barrier
between the two wells in one subsystem and together with the width $\sigma$ of
the barrier it relates to the coupling strength $v$ in the matrix model.  Again
the coupling between the two subsystems is controlled by a parameter labelled
$\gamma$.

To solve this equation we use an ansatz of coupled Gaussian functions (compare
\cite{Rau10a,Rau10b}),
\begin{equation}
	\psi
		= \sum_{i=A,B \atop j=1,2} \psi_{i,j}
		= \sum_{i=A,B \atop j=1,2} \exp\left({a_{i, j} x^2 + b_{i,j} x + c_{i,j}}\right).
	\label{eq:gauss_ansatz}
\end{equation}
We use four wave functions, two for each subsystem ($i=A,B$) and place one in
each well ($j=1,2$). Again we place restrictions on our ansatz. We require that
the norm of each subsystem is one, which reduces our parameter set by two. In
addition we require
\begin{eqnarray}
	\Im c_{A, 1} = \varphi_A, \qquad&
	\Im c_{B, 1} =  \varphi_{\rm B} + \varphi_{\rm rel}, \nonumber\\
	\Im c_{A, 2} = -\varphi_A, \qquad&
	\Im c_{B, 2} = -\varphi_{\rm B} + \varphi_{\rm rel}
\end{eqnarray}
with a constant $\varphi_{\rm rel}$ determining the phase difference between
the two modes, and again reducing the parameter set by two. Therefore from the
24 parameters $a_{i,j},b_{i,j},c_{i,j} \in \mathbb{C}$ 20 free parameters
remain and must be determined such that adequate solutions are found. With
these constraints the ansatz is consistent with the ansatz for the matrix model
and the system with the double-$\delta$ potential.

To obtain solutions of \eref{eq:gauss} we apply the time-dependent variational
principle \cite{McLachlan1964a} to the time-dependent GPEs
\begin{eqnarray}
	\Big( -\frac{\partial^2}{\partial x^2}
	- g | \psi_{\rm A} |^2
	+ \frac{1}{4}	x^2 + V_0^{\rm G} \ee^{-\sigma x^2} &\Big) \psi_{\rm A}
	+ \ii \gamma x \ee^{-\rho x^2} \psi_{\rm B}
	&= \ii \frac{\partial}{\partial t} \psi_{\rm A},
	\nonumber\\
		\Big( -\frac{\partial^2}{\partial x^2}
	- \underbrace{g | \psi_{\rm B} |^2}_{\rm contact}
	+ \underbrace{\frac{1}{4} x^2 + V_0^{\rm G} \ee^{-\sigma x^2}}_{\rm trap}
			&\Big)  \psi_{\rm B}
	- \underbrace{\ii \gamma x \ee^{-\rho x^2} \psi_{\rm A}}_{\rm coupling}
	&= \ii \frac{\partial}{\partial t} \psi_{\rm B}. \label{eq:gauss_tdvp}
\end{eqnarray}
We search a parameter set for our ansatz, which minimizes the difference
between the left-hand and right-hand side of the equation, viz.\ we determine
the minimum of the functional
\begin{eqnarray}
	I = \left\Vert H \psi - \ii \phi \right\Vert^2.
\end{eqnarray}
In this procedure $\psi(t)$ is kept constant for a given point in time and
$\dot \psi = \phi$ is varied to minimize $I$. Since the wave function
$\psi(z(t))$ is not varied we require that the parameters $z = \{a_{i,j},
b_{i,j}, c_{i,j}\}$ do not change. A variation with respect to $\dot z$ leads
to the equations of motion for the variational parameters, which follow from
\begin{eqnarray}
	\left< \frac{\partial\psi}{\partial z} \middle| \dot \psi -
		\ii H \psi\right> = 0.
\end{eqnarray}
A more elaborate explanation of the method can be found in \cite{Dast13}. With
a numerical root search we can now determine those states which satisfy the 20
conditions
\begin{eqnarray}
	0 = \dot a_{i, j}, 0 = \dot b_{i,j}, \\
	\mu_i = \ii \dot c_{i,1}^* = \ii \dot c_{i,2}^*
	\Rightarrow 0 = \dot c_{i,1} - \dot c_{i,2} ~{\rm with}~ i = A,B.
\end{eqnarray}
For \PT-symmetric solutions the chemical potentials of the subsystems will
fulfil $\mu_A = \mu_B \in \mathbb{R}$.

\section{\texorpdfstring{\PT}{PT}-symmetric properties and
	bifurcation structure of the systems}
\label{sec:3}

First we will examine analytical solutions of the matrix model. The bifurcation
structure of these solutions and their \PT-symmetric properties will be
discussed. Furthermore the differences and similarities between this
four-dimensional hermitian matrix model and the two-dimensional matrix model
with imaginary potential will be examined.

In a next step the results obtained with the matrix model will be compared 
with the spatially extended models. Also the influence of the phase difference
between the two modes will be investigated.

\subsection{Bifurcations structure and
	\texorpdfstring{\PT}{PT}-symmetric properties of the matrix model}

To obtain analytical solutions we have to impose some constraints on the ansatz
of the wave function of the matrix model \eref{eq:complexmu}. \PT-symmetric
solutions must fulfil the condition \eref{eq:ptsymmetrycondition} which for
this matrix model results in
\begin{eqnarray}
	\psi_{j, 1} &= \psi_{j, 2}^*
	\quad{\rm with}~
	j=A, B
	\label{eq:ptcondition2}
\end{eqnarray}
and
\begin{eqnarray}
	\psi_{A, i} &= \psi_{B, i}^*
	\quad{\rm with}~
	i=1, 2.
	\label{eq:ptcondition3}
\end{eqnarray}
This ensures that the particle loss in one system is compensated by the other.
These restrictions lead to the ansatz
\begin{eqnarray}
	\psi = \frac{1}{\sqrt{2}} \left( \ee^{\ii\varphi}, \ee^{-\ii\varphi},
			\ee^{-\ii\varphi}, \ee^{\ii\varphi} \right)
	\label{eq:ptsymansatz}
\end{eqnarray}
with which we obtain an analytical expression for the chemical potentials of two
\PT-symmetric solutions
\begin{eqnarray}
	\mu = -\frac{g}{2}\pm \sqrt{v^2 + \gamma^2}.
	\label{eq:analytic_sym}
\end{eqnarray}
A more detailed calculation is given in \ref{s:analyticalsolutions}.

\begin{figure}
  \noindent\includegraphics[width=0.99\textwidth]{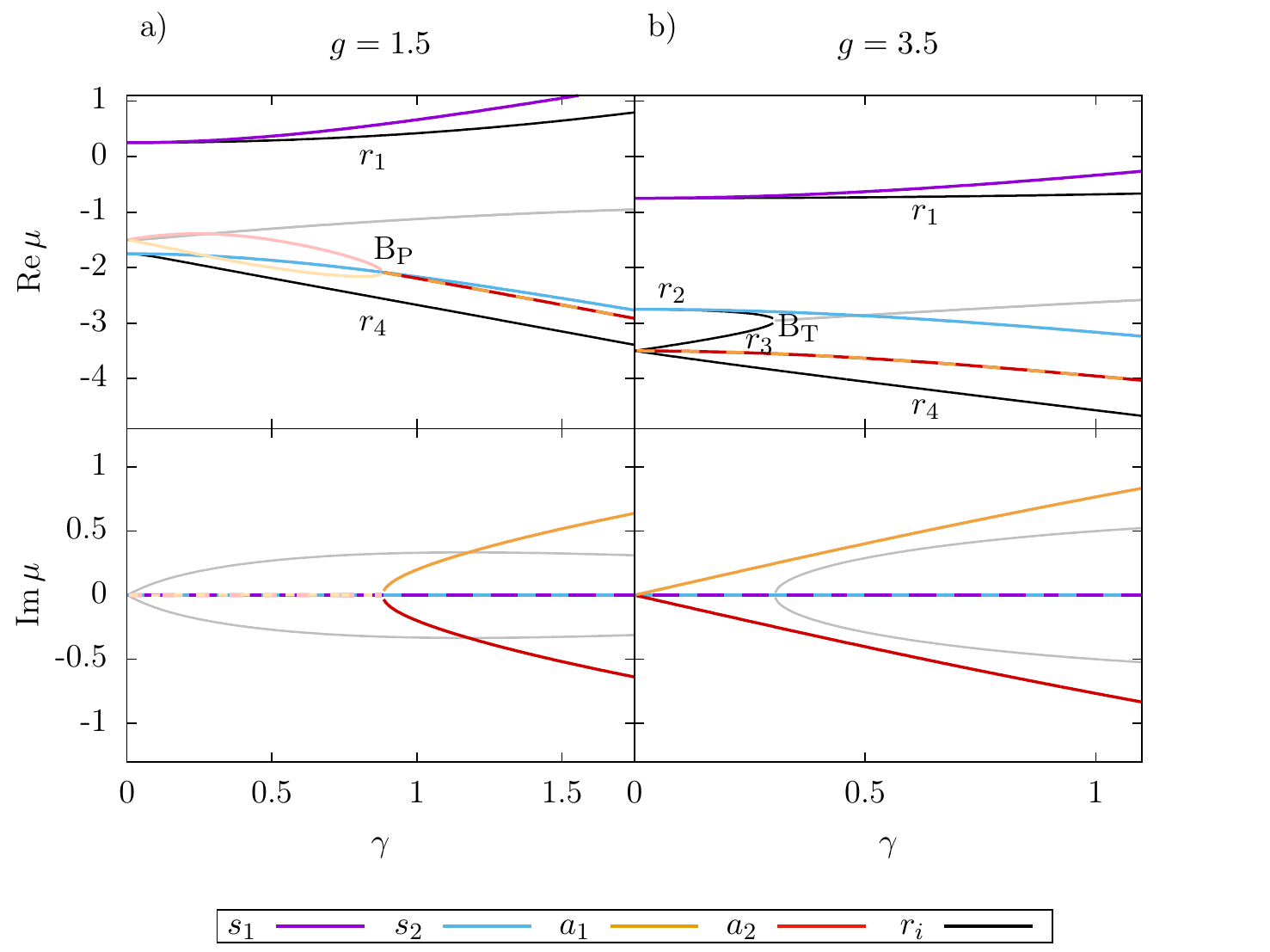}
	\caption{ Analytical solutions for the chemical potential of \eref{eq:matrix}
		are shown. The \PT-symmetric states are denoted by $s_1$ and $s_2$.
		\PT-broken states are labelled with $a_1$ and $a_2$.  Solutions of the
		effective system \eref{eq:ef2dmatrix} are labelled with $r_i$. The states
		$r_2$ and $r_3$ only exist for $|g| > 2$ and therefore appear only in
		figure b. The coupling strength is set to $v=1$. In a) the nonlinearity is
		set to $g = 1.5$ while in b) it is set to $g = 3.5$. The pitchfork
		bifurcation between the states $a_{1,2}$ and $s_2$ in a) is labelled with
		$\rm B_P$ and occurs at $\gamma \approx 0.882$.  The tangent bifurcation
		between states $r_2$ and $r_3$ in b) is marked by $\rm B_T$. The
		analytically continued solutions are plotted using lighter colours.}
	\label{fig:matrix}
\end{figure}

The solutions are plotted in \fref{fig:matrix} and labelled $s_1$ and $s_2$.
For different values of $g$ the solutions are shifted up or down. For
increasing values of $\gamma$ the difference of the values of the chemical
potential of the two states is increased.

\PT-broken states do not need to obey condition \eref{eq:ptcondition2} but
\eref{eq:ptcondition3} still must be fulfilled since the influx and outflux
between subsystem A and B must be equal. Therefore the ansatz for these states
reads
\begin{eqnarray}
	\psi = \left( \cos\theta \ee^{\ii\varphi}, \sin\theta \ee^{-\ii\varphi},
					\cos\theta \ee^{-\ii\varphi}, \sin\theta \ee^{\ii\varphi} \right)
	\label{eq:ptbrokenansatz}
\end{eqnarray}
with $\mu_A = \mu_B^*$ . The calculation in \ref{s:analyticalsolutions} yields
the analytical expressions for the chemical potentials
\begin{eqnarray}
	\mu_{\rm A} = \mu_{\rm B}^*
		= -\frac{g}{2} \left( 2 \mpsb \sqrt{P+\frac{\gamma^2}{v^2}P^2}-P \right)
		~{\rm with}~
	P = \frac{1}{2} \pmsa \frac{\sqrt{g^2 + 16 \gamma^2}}{2g}.
	\label{eq:analytic_asym}
\end{eqnarray}
Note that the $\mpsb$ and $\pmsa$ are independent and we therefore obtain four
expressions \eref{eq:analytic_asym} for \PT-broken states. However two of these
solutions only exist in an analytically continued system (see \fref{fig:matrix}).

For $|g| < 2v$ the state $s_2$ passes through a pitchfork bifurcation at
\begin{eqnarray}
	\gamma_c = \sqrt{\frac{4v^4}{g^2} - v^2},
\end{eqnarray}
in which $a_1$ and $a_2$ are created. For $\gamma > \gamma_c$ these two states
have the same $\Re \mu_{\rm A}$ but a complex conjugate $\Im \mu_{\rm A}$. This
means that one of the states gains particles in subsystem A while in subsystem
B it is depleted, and vice versa. The pitchfork bifurcation occurs at smaller
values of $\gamma_c$ for an increasing nonlinearity $g$ until for $|g| = 2v$
the value of $\gamma_c$ reaches zero. For values of $|g| > 2v$ the bifurcation
between $a_{1,2}$ and $s_2$ no longer occurs and the states $a_{1,2}$ exist
independent of $s_2$ for all $\gamma$. For $g < 0$ the bifurcation occurs not
with the state $s_2$ but with $s_1$. Thus we have shown that \PT-symmetric
states exist for the closed four-dimensional hermitian matrix model and
\PT-symmetry breaking can be observed. 

Besides these states there is another class of states in the four-dimensional
matrix model. Wave functions which fulfil the condition
\begin{eqnarray}
	\psi_{{\rm A},i} = -\ii\psi_{{\rm B},i}
	~{\rm with}~
	i = 1,2
	~{\rm and}~
	\psi_{{\rm A},i}, \ii\psi_{{\rm B},i}\in \mathbb{R}
\end{eqnarray}
lead to decoupled equations for $\psi_A$ and $\psi_B$ and result in the
effective two-dimensional model
\begin{eqnarray}
	\left( \begin{array}{cc}
    -g | \psi_1 |^2 - \gamma & v \\
		v & -g | \psi_2 |^2 + \gamma
  \end{array} \right) 
  \left( \begin{array}{c} \psi_1 \\ \psi_2 \end{array} \right)
	=
 	\mu \left( \begin{array}{c} \psi_1 \\	\psi_2 \end{array} \right)
	~{\rm and}~
	\psi_{1,2} \in \mathbb{R}.
	\label{eq:ef2dmatrix}
\end{eqnarray}
These states effectively describe a double-well system with a real potential,
where one potential well is lowered and the other is raised by the value of
$\gamma$.  As expected we find that the amplitude of the wave function in the
higher well is lower than in the other.  For $\gamma = 0$ the system crosses
over to a symmetric double-well model with no coupling and therefore we can see
in \fref{fig:matrix}a that the state $r_1$ merges with the state $s_1$ and the
state $r_4$ merges with the state $s_2$. For values $g > 2v$ the bifurcation
between the \PT-symmetric and \PT-broken states no longer exists and two new
states $r_2$ and $r_3$ emerge. Now at $\gamma = 0$ the states $r_1$ and $s_1$
as well as $r_2$ and $s_2$ become equal. Also $r_3$, $r_4$ and $s_4$ merge. For
increasing $\gamma$ the states $r_2$ and $r_3$ vanish in a tangent bifurcation.
The method used to solve \eref{eq:ef2dmatrix} is described in
\ref{s:analyticalsolutions}.

We can compare the results of the four-dimensional matrix model in
\fref{fig:matrix} with those of the two-dimensional matrix model shown in
\fref{fig:2dmatrix}. It is immediately clear that our system shows a new and
richer bifurcation scenario which differs from the two-dimensional matrix
model.  While the \PT-symmetric eigenvalues of the states in the
two-dimensional system approach each other for increasing coupling strengths
$\gamma$ until they merge in a tangent bifurcation, in our system the
eigenvalues increase in distance for larger values of $\gamma$ and no
bifurcation between the two states $s_{1,2}$ occurs. However, some generic
features remain the same. In both cases the \PT-symmetric state $s_2$ with a
real $\mu$ passes through a pitchfork bifurcation, out of which the \PT-broken
states with complex $\mu$ emerge. For both models this bifurcation moves to
smaller values of $\gamma$ until, for a critical value of the nonlinearity $g$,
the bifurcation vanishes and the \PT-symmetric and \PT-broken states never
coincide.

One advantage of using a matrix model compared to systems with a more realistic
spatially extended description is that the matrix model gives an overview over
all possible effects in a system while remaining straightforward to calculate.
Also the knowledge about symmetry properties and existence of states gained
from the matrix model can help finding states in the more complicated models,
e.g.\ by choosing appropriate initial values for a root search.  Since we want
to concentrate our investigation on the \PT-symmetric properties of the
subsystems, we will not further investigate the states $r_i$.

\subsection{Comparison of the matrix model and the model with a spatial
	resolution of the wave function}

\begin{figure}
  \noindent\includegraphics[width=0.99\textwidth]{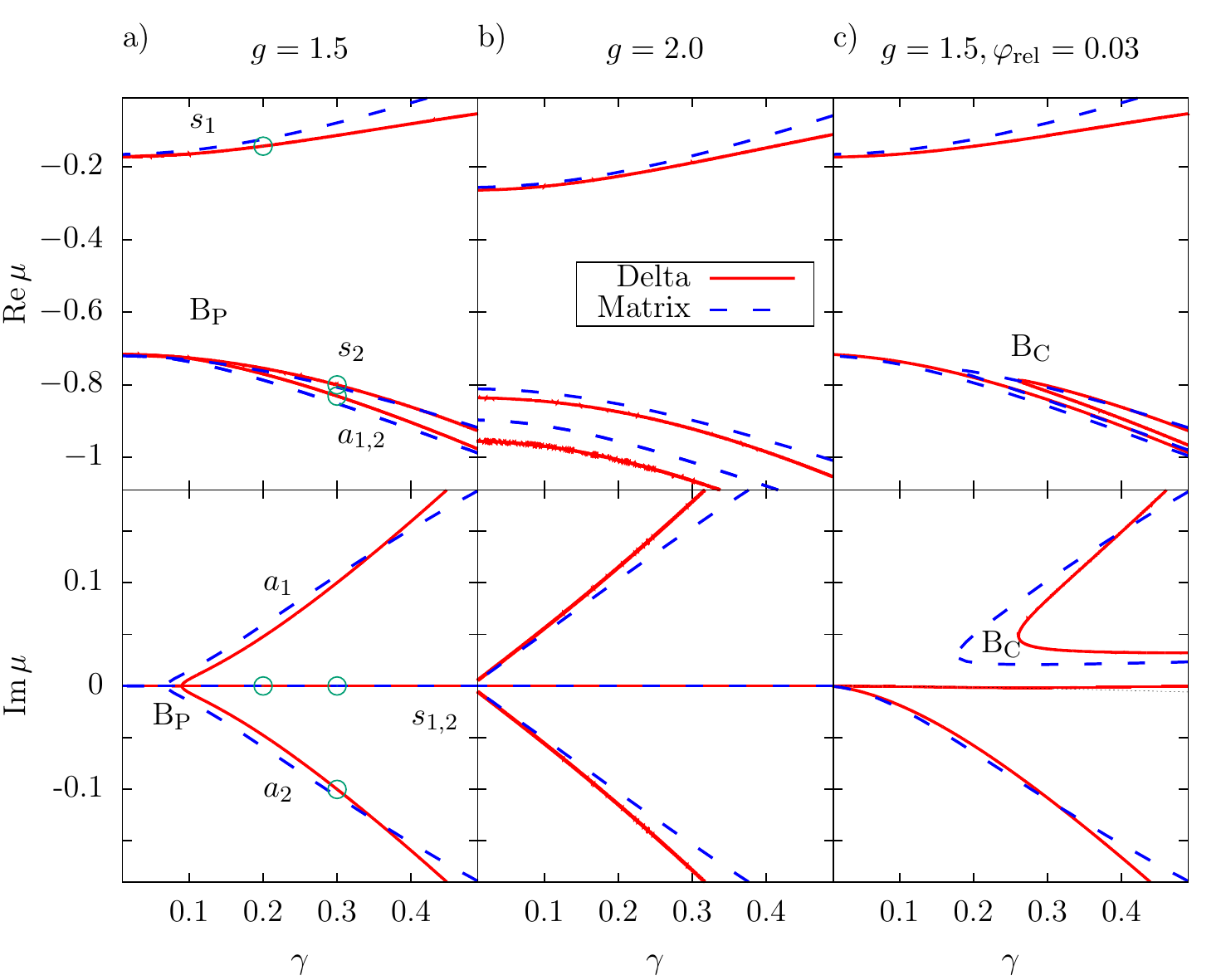}
	\caption{ Chemical potential $\mu = \mu_{\rm A} = \mu_{\rm B}^*$ for the
		matrix model \eref{eq:complexmu} (blue dashed lines).  The parameters of the
		matrix used for all three plots are $g_0 = 2.75$, $v = 0.28$ and $\gamma_0
		= 1.27$. The shift of the chemical potential of the matrix model is
		$\Delta\mu = -0.17$.  For both figures a) and b) the phase difference
		$\varphi_{\rm rel}$ was set to zero. Figure a) was calculated for a
		nonlinearity of $g = 1.5$.  The different states are denoted by
		$s_{1,2}$, $a_{1,2}$. In plot b) a nonlinearity of $g = 2.0$ was used. For
		plot c) the same nonlinearity as in plot a) was used but the phase
		difference was set to $\varphi_{\rm rel} = 0.03$.  The figure also contains
		the results for the double-$\delta$-system (red solid lines). For the coupling
		of the two subsystems $V_0^{\rm D}$ was set to $1.0$ and the
		$\delta$-potentials were located at $b = \pm1.1$. The same nonlinearities
		as for the matrix model were used.  In figure a) the parameters for which
		the wave functions are shown in \fref{fig:deltawave} are marked by green
		circles. A pitchfork bifurcation between the states $s_2$ and $a_{1,2}$ is
		denoted by $\rm B_P$. An additional cusp bifurcation appearing in the case
		$\varphi_{\rm rel}$ is marked by $\rm B_C$.}
	\label{fig:deltamatrix}
\end{figure}

\begin{figure}
  \noindent\includegraphics[width=0.99\textwidth]{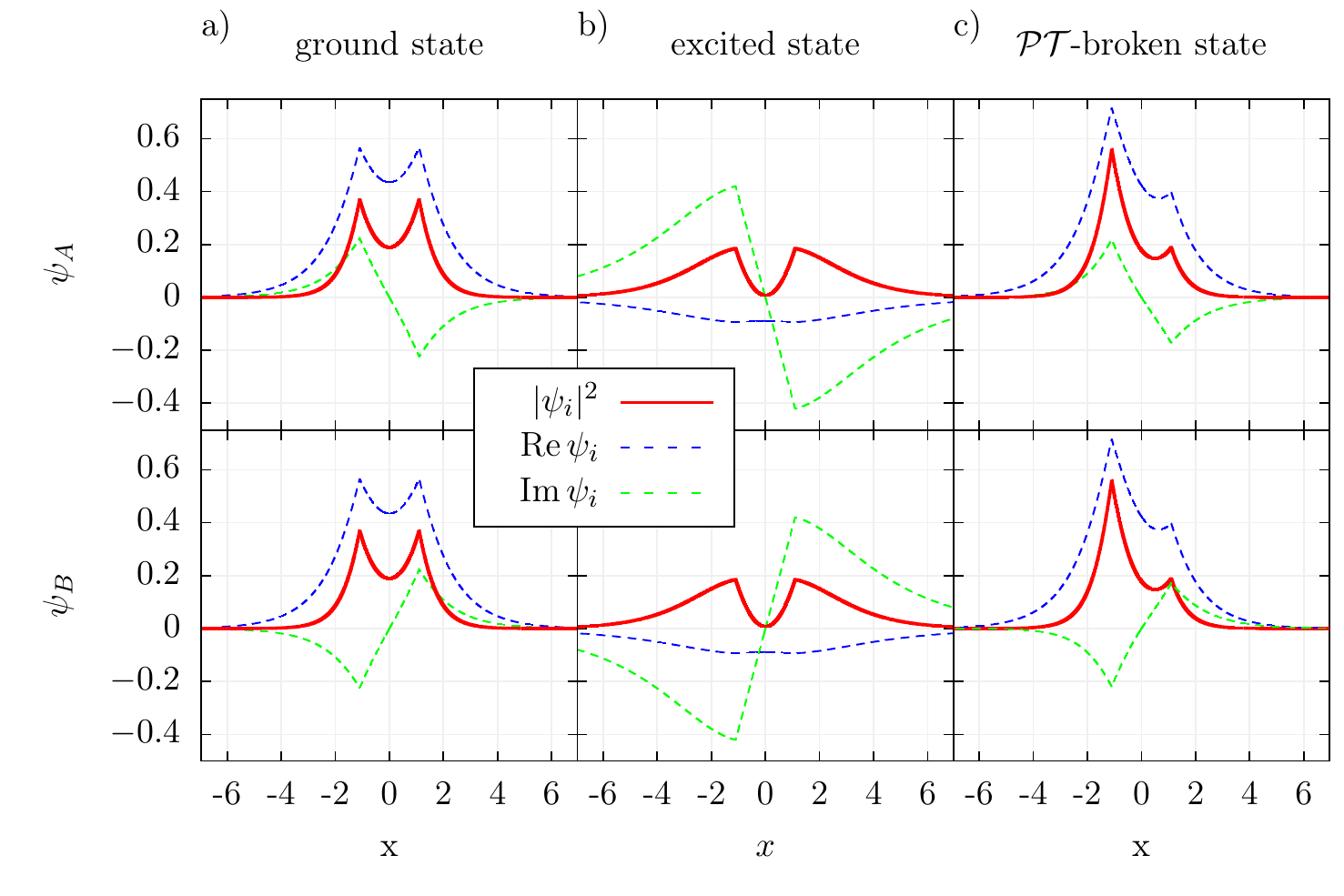}
	\caption{ Wave functions of the double-$\delta$ potential system for the
	parameter sets marked in \fref{fig:deltamatrix}a.  a) Wave function of the
	\PT-symmetric ground state.  b) Wave function of the \PT-symmetric excited
	state.  In c) the broken symmetry of the \PT-broken state can be recognized.}
	\label{fig:deltawave}
\end{figure}

The results of the system with the double-$\delta$ potentials are given in
\fref{fig:deltamatrix} in comparison with those of the matrix model.  To be able
to compare the two models the parameters in the matrix model are replaced by $g
\rightarrow {g}/{g_0}$ and $\gamma \rightarrow {\gamma}/{\gamma_0}$. Also a
shift $\Delta\mu$ in the chemical potential is introduced. Then the parameters
$\gamma_0$, $g_0$, $v$ and $\Delta\mu$ are fitted to the results of the
double-$\delta$ model. How these parameters are connected to the extended model
can be seen in \ref{s:matrixmodelderivation}.

In contrast to the matrix model the double-$\delta$ system includes spatial
properties of the wave functions. In \fref{fig:deltawave} the wave functions for
the parameters marked in \fref{fig:deltamatrix} are shown. One can clearly
observe the non-differentiability of the wave functions at the locations $x =
\pm b$ of the $\delta$-potentials. It is also clearly visible that the states
with complex chemical potential are \PT-broken (see \fref{fig:deltawave}c). The
two wave functions for the subsystems A and B fulfil the condition $\psi_{\rm
A}(x) = \psi^*_{\rm B}(x)$ which ensures that the loss and gain in each
subsystem is balanced by the gain and loss in the other subsystem and the
\PT-symmetry of the potential is maintained.  Furthermore the wave function of
the ground state (\fref{fig:deltawave}a) is much more localized in the potential
wells than the wave function of the excited state (\fref{fig:deltawave}b).

When we compare the solutions of the matrix model with those of the model with
the double-$\delta$ potential we observe that the qualitative bifurcation
structure of the states is the same for both models but some quantitative
deviations can be seen. Before we continue our investigation of the cause of
these differences in \sref{sec:3.3} we will take a look at the influence of the
phase difference $\varphi_{\rm rel}$ between the two subsystems.

To examine the influence of the phase difference on the bifurcation scenario
we show in \fref{fig:deltamatrix}c the case in which the phase difference
between the subsystems is set to $\varphi_{\rm rel} = 0.03$. The pitchfork
bifurcation ${\rm B}_{\rm P}$ in \fref{fig:deltamatrix}c turns into a cusp
bifurcation ${\rm B}_{\rm C}$. While the central (\PT-symmetric) state $s_1$
exists on both sides of the bifurcation point, the two outer (\PT-broken)
states $a_{1,2}$ are created in the bifurcation of \fref{fig:deltamatrix}a. In
the cusp bifurcation of \fref{fig:deltamatrix}c one of the outer states
(depending on the sign of $\varphi_{\rm rel}$) merges with the central state
and the other outer state performs a continuous transition to the central
state for smaller values of $\gamma$.  Also the \PT-symmetry of all states is
broken. The asymmetry increases for the central state for increasing values of
$\varphi_{\rm rel}$.

If we introduce the phase difference $\exp(\ii \varphi_{\rm rel})$ between the
to subsystems explicitly into the stationary GPE \eref{eq:matrixansatz} for
the matrix model, we obtain for the subsystem A
\begin{eqnarray}
	\mu_A \psi_{\rm A, 1} &= - g | \psi_{\rm A, 1} |^2 \psi_{\rm A, 2}
		+ v \psi_{\rm A, 2}
		+ \sin(\varphi_{\rm rel})     \gamma \psi_{\rm B, 1}
		- \ii \cos(\varphi_{\rm rel}) \gamma \psi_{\rm B, 1}, \nonumber \\
	\mu_A \psi_{\rm A, 1} &= - g | \psi_{\rm A, 1} |^2 \psi_{\rm A, 2} 
		+ v \psi_{\rm A, 1}
		- \underbrace{\sin(\varphi_{\rm rel})     \gamma \psi_{\rm B, 2}}_{\rm asym.\ pot.}
		+ \underbrace{\ii \cos(\varphi_{\rm rel}) \gamma \psi_{\rm B, 2}}_{\rm gain\ or\ loss}.
\end{eqnarray}
We see that a phase difference between the two subsystems leads to different
contributions to the real and imaginary part of the effective potential of
each subsystem. The real part of the effective potential can therefore become
asymmetric (this not only depends on the phase difference $\varphi_{\rm
rel}$ but also on the phase value of the wave function in the other subsystem).

The influence of an asymmetric double-well potential on the bifurcation
structure has been discussed previously \cite{PhysRevE.74.056608}. For
an asymmetric potential there is no longer a pitchfork bifurcation but a
tangent bifurcation. We can compare this to the well known normal forms of the
two parameter bifurcation theory \cite{Kuznetsov2004}. The normal form
of the cusp bifurcation is
\begin{eqnarray}
	0 = \dot x = f_{\rm C}(x) = \beta + \alpha x - x^3, \label{eq:normalcusp}
\end{eqnarray}
with the bifurcation parameters $\alpha$ and $\beta$. In our model the role of
the second parameter $\beta$ is taken by the phase difference $\varphi_{\rm
rel}$ between the two subsystems. A constant $\varphi_{\rm rel} = 0$ (which is
equivalent with $\beta = 0$) defines a line in the $\varphi_{\rm rel}$-$\gamma$
parameter space. On this line the pitchfork bifurcation scenario emerges.

We have seen that the phase difference between the two modes is critical to
obtain a \PT-symmetric system, and the breaking of this symmetry
changes the bifurcation structure. Only for $\varphi_{\rm rel} = 0$
\PT-symmetric states are observed.

\subsection{Comparison of the models and usefulness of the matrix model}
\label{sec:3.3}

\begin{figure}[t]
  \noindent\includegraphics[width=0.90\textwidth]{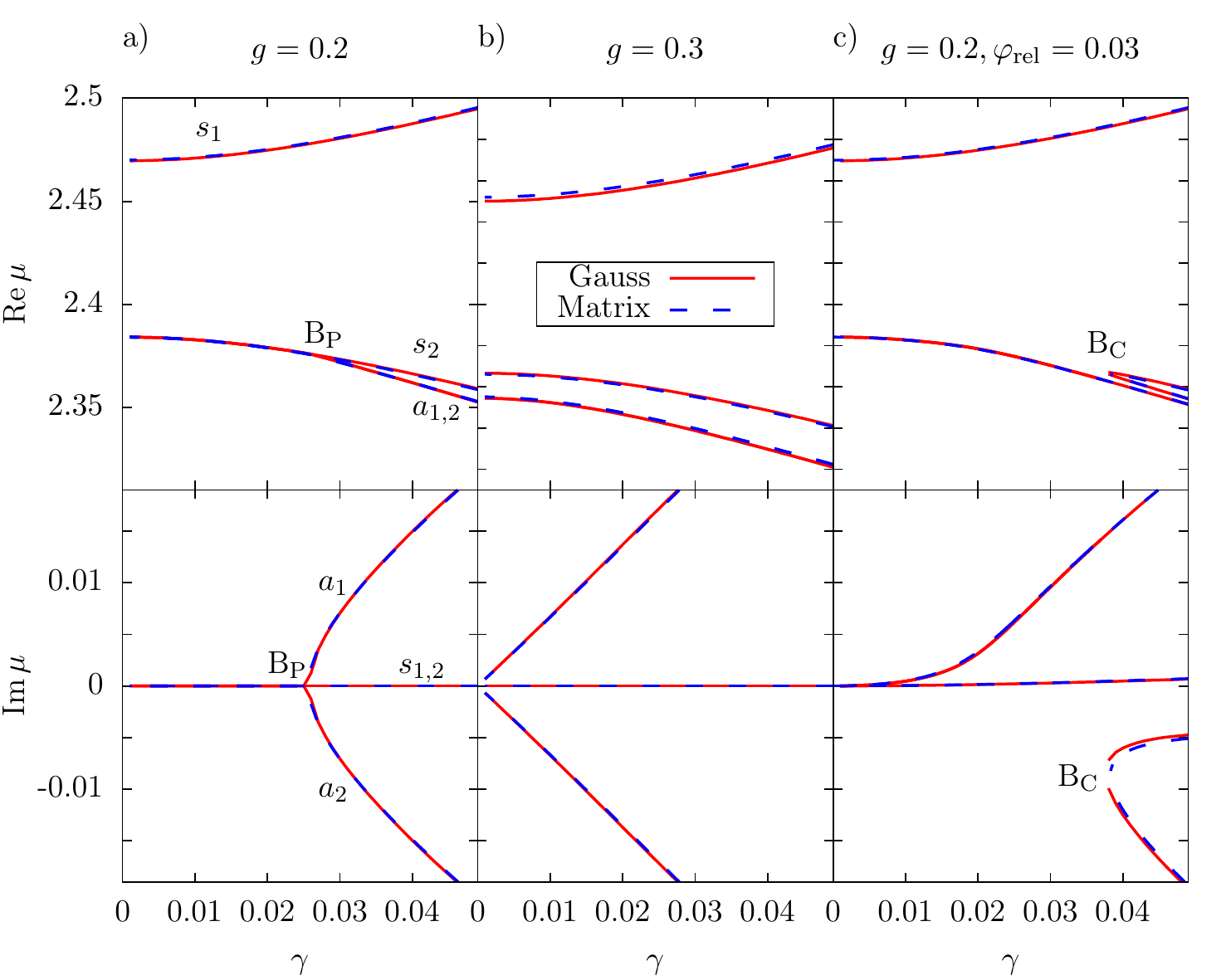}
	\caption{ Comparison of the eigenvalues of the matrix model from 
		\eref{eq:matrix} (blue dashed lines) with the eigenvalues of the system
		\eref{eq:gauss} (red solid lines), in which the BEC is trapped in a smooth
		harmonic potential separated into two wells by a Gaussian potential
		barrier. The fit parameters for the matrix model are $g_0 = 2.78$, $v =
		0.043$ and $\gamma_0 = 0.92$ and are used for all cases a)-c). The chemical
		potential of the matrix model is shifted by $\Delta\mu =2.463$. The height
		of the Gaussian potential barrier in system \eref{eq:gauss} is $V_0^{\rm G}
		= 0.25$ with the width $\sigma = 0.5$. Figures a) and c) contain the results
		for $g = 0.2$, while figure b) is plotted for $g = 0.3$. In figure c) the
		phase difference is non-zero ($\varphi_{\rm rel}=0.03$). }
	\label{fig:gaussmatrix}
\end{figure}

In the system \eref{eq:gauss} the two modes are coupled over a spatially
extended range and therefore the continuous change of the phase in the wave
functions may play a role. In \fref{fig:gaussmatrix} we show the stationary
states of the matrix model \eref{eq:matrix} in comparison with those of the
smooth potential system \eref{eq:gauss}. The parameters of the matrix model
($g_0$,$\gamma_0$ and $v$) and a shift of the chemical potential $\Delta\mu$
were adjusted to the solution of the model \eref{eq:gauss} but remained the
same for all calculations in \fref{fig:gaussmatrix} with different values for
$g$ and $\varphi_{\rm rel}$. In \ref{s:matrixmodelderivation} it is shown how
the discrete matrix model can be derived from a continuous model.

Again we see a pitchfork bifurcation (\fref{fig:gaussmatrix}a) in the lower
state which, for increasing values of the nonlinearity $g$, moves to smaller
values of $\gamma$. The two new states created in this bifurcation are
non-stationary ($\mu_{\rm A, B} \not\in \mathbb{R}$) \PT-broken states. By
further increasing $g$ the value of $\gamma$ at which the bifurcation occurs
moves to even smaller values of $\gamma$ until it reaches $\gamma = 0$.  Thus
the qualitative behaviour is exactly the same as in the two previously
investigated models. It is generic for the coupled double-well structure. If
the phase between the two subsystems is changed to a non-zero value, the
pitchfork bifurcation from \fref{fig:gaussmatrix}a changes into a cusp
bifurcation (compare \fref{fig:gaussmatrix}c). This is the same behaviour as
observed in \fref{fig:deltamatrix} for the double-$\delta$-potential. No change
of the bifurcation structure or the \PT-symmetric properties due to the
extended coupling is observed. However, as can be seen in
\fref{fig:gaussmatrix}a-c the agreement with the matrix model is nearly perfect
and much better than the agreement between the matrix model and the model with
the $\delta$-potential wells.

\begin{figure}
  \noindent\includegraphics[width=0.95\textwidth]{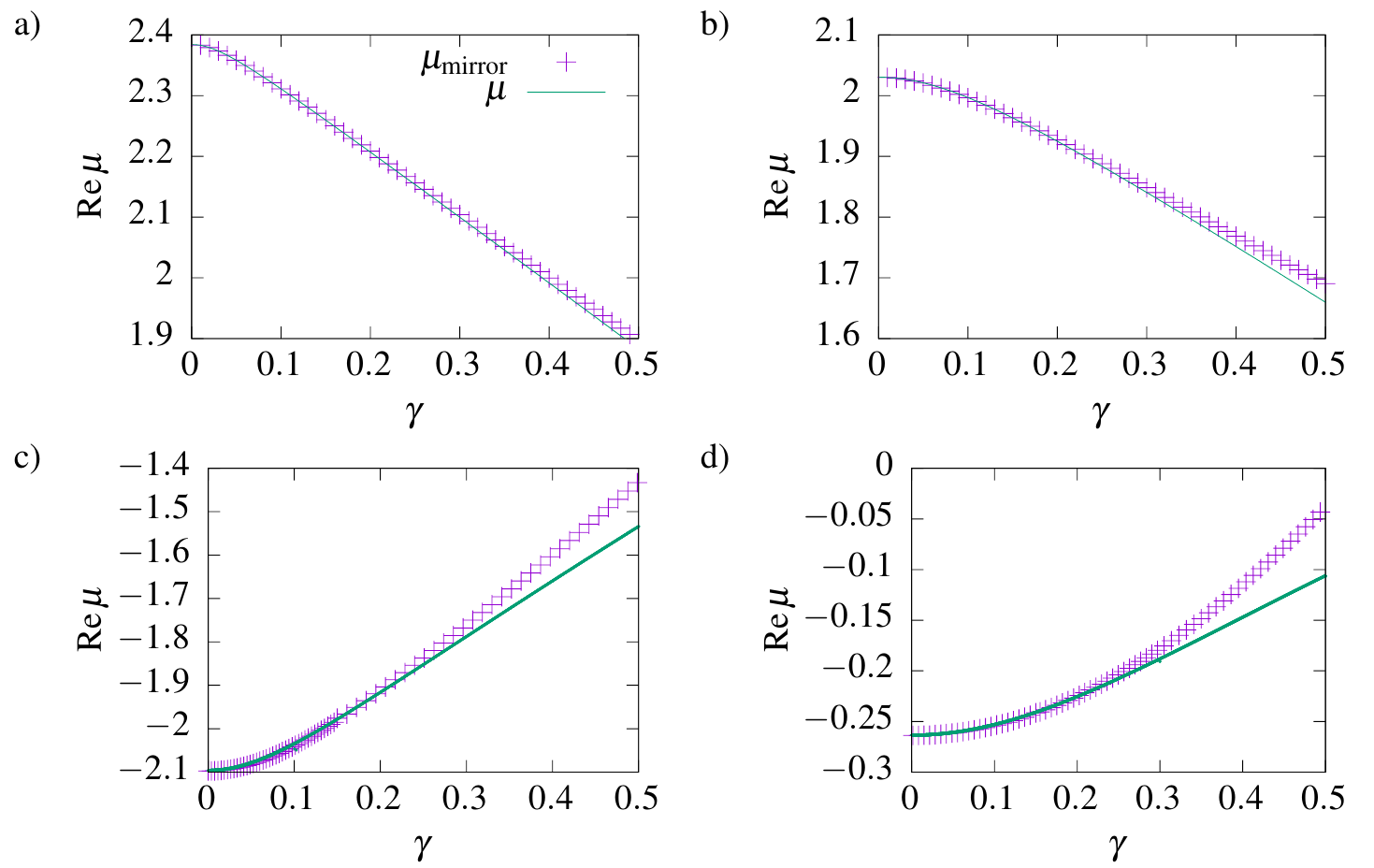}
	\caption{ Ground state and mirrored excited state ($\mu_{\rm mirror} = \mu_0
		- \mu$). The states are not symmetric. Figures a) and b) show the results
		for the Gaussian model \eref{eq:gauss} with $g = 0.2$ and $\mu_0 = 4.854$
		and $\mu_0 = 4.2733$, respectively. Figures c) and d) show the results of
		the double-$\delta$ model \eref{eq:deltaGPEb} with $g = 2.0$ and $\mu_0 =
		-4.5$ and $\mu_0=-1.1$, respectively. In the Gaussian model the hight of
		the potential barrier between the two wells in each subsystem is changed.
		For a) the barrier hight is $V_0^{\rm G} = 4.0$, for b) it is $V_0^{\rm G}
		= 2.5$. In the case of the $\delta$-model the (real) depth of the
		potentials is lowered from $V_0^{\rm D} = 1.0$ in a) to $V_0^{\rm G} = 2.5$
		in b).}
	\label{fig:locality}
\end{figure}

Taking a closer look at the states of the matrix model one discovers that the
upper and lower states are symmetric with respect to $-{g}/{2}$ as can be seen
in \eref{eq:analytic_sym}. This is no longer true for the models with a spatial
description. To make this asymmetry visible we examine \fref{fig:locality} in
which one state is mirrored onto the other, e.g. for one state
\begin{eqnarray}
	\mu_{\rm mirror} = \mu_0 - \mu
\end{eqnarray}
is plotted and $\mu_0$ is the average value of the chemical potential of both
states at $\gamma = 0$. One observes that the deviation is much more pronounced
in the model with $\delta$-wells than in the smooth potential from
\eref{eq:gauss}.

\begin{table}[t]
	\caption{ \label{tab:parameter}
	   Fit parameters of the matrix model used for the comparison with the
		 spatially extended models in figures \ref{fig:deltamatrix} and
		 \ref{fig:gaussmatrix}.}\lineup
	\begin{indented}
		\item[] \begin{tabular}{@{}l|llll|ll|ll}
			\br
			Comparison with &
			$ g_0 $ & $ v $ & $\gamma_0$ & $\Delta\mu$ & $ V_0^{\rm G} $ & $\sigma$ & $ V_0^{\rm D} $ & $b$ \\
			\mr
			double-$\delta$ model &
			$2.75$ & $0.28$ & $1.27$  & $-0.17$ & --- & --- & $1.0$ & $1.1$
			\\ 
			smooth potential &
			$2.78$ & $0.043$ & $0.92$ & \m$2.463$ & $2.5$ & $0.5$ & --- & ---
			\\
			\br
		\end{tabular}
	\end{indented}
\end{table}

In the comparison of the fit parameters $g_0$, $v$ and $\gamma_0$ (see
\tref{tab:parameter}), one parameter with vastly different values is evident.
The coupling strength $v$ of the two potential wells in the $\delta$-potential
model case is approximately $6.5$ times larger than in the case of the harmonic
trap with the potential well.  This means that the separation of the two wells
is much less pronounced due to shallower wells in the case of the
$\delta$-potential. This leads to wave functions which are not as localized as
in the case of the smooth potential. Therefore the contribution of the overlap
of the wave functions, which was negligible  for the smooth potential,
increases. The matrix model is not capable of describing the nonlinear
interaction between wave functions of different modes. Only the nonlinear
scattering process in the same well is taken into account.

For further investigation one can increase the distance between the wells or
deepen them. One might expect that the stationary states then would be in a
better agreement with the matrix model. We compare the model with smooth
potentials for different barrier heights (\fref{fig:locality}a and
\fref{fig:locality}b). For a lower potential barrier the asymmetry of the two
states becomes more pronounced. The same is true for the $\delta$-model
(\fref{fig:locality}c and \fref{fig:locality}d).

\begin{figure}
  \noindent\includegraphics[width=0.95\textwidth]{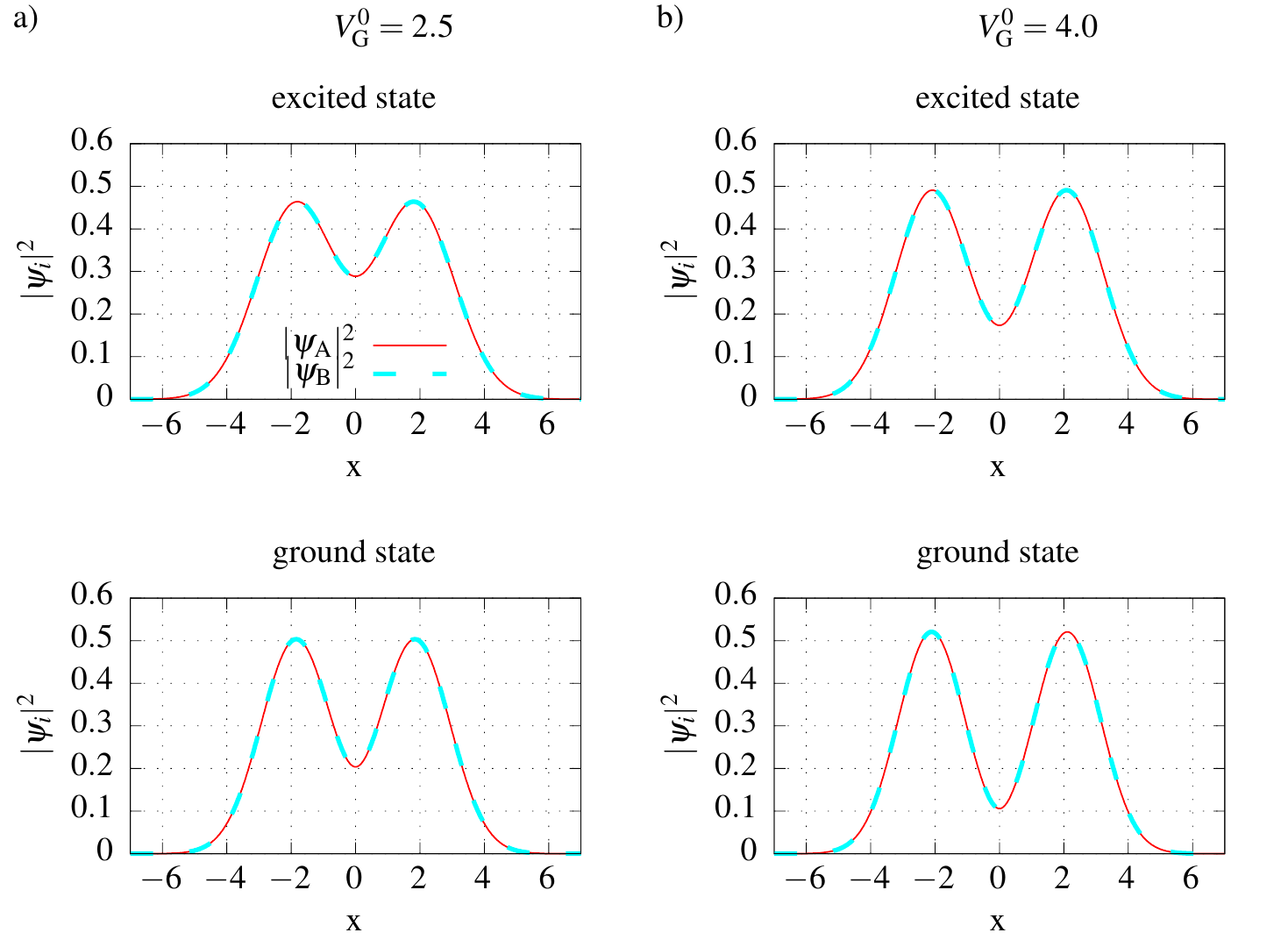}
	\caption{ Wave functions for the ground and excited	states in the Gaussian
		model for different potential barriers (in a) $V_0^{\rm G} = 2.5$, in b)
		$V_0^{\rm G} = 4.0$) for a nonlinearity of $g = 0.2$. The overlap of the
		Gaussians at $x = 0$ is much higher for the lower potential barrier in a)
		and for the excited states. }
	\label{fig:wavefct}
\end{figure}

The wave functions for the different parameter sets are shown in
\fref{fig:wavefct}. Here the probability density of the ground and excited
state for the smooth potential model with different heights for the potential
barrier can be seen. One observes a higher probability density in the overlap
region around $x=0$ for the excited states.  This overlap increases for a lower
potential barrier. Thus, we can conclude that the matrix model captures all
relevant information of the bifurcation scenario and the \PT-symmetric
properties as long as the different potential wells are sufficiently separated.
A larger overlap leads to quantitative changes and the loss of a mirror
symmetry of pairs of values for the chemical potential in the ($\mu$,
$\gamma$)-diagram, however, it does not affect the generic structure of the
states.

\section{Summary and Outlook}
\label{sec:4}

For an experimental realization of a \PT-symmetric double-well potential the
description of a physical environment which implements the gain and loss of a
complex potential is an important prerequisite. By combining two double-well
subsystems into one closed hermitian system we have found such a realization.

For the four-dimensional matrix model without a phase difference between the
two subsystems analytical solutions for all \PT-symmetric and \PT-broken
states were found. Although the four-dimensional matrix model showed a new
and different bifurcation scenario in comparison with the two-dimensional
matrix model from \cite{Graefe12} some generic features remained the same.

The matrix model showed the same qualitative bifurcation scenario as the two
spatially extended models. Deviations could be observed when the two wells of
the systems were not isolated enough such that the wave functions in each well
had a significant overlap between the wells. In this case the solutions from
the systems with a spatially resolved wave function differed from those of the
matrix model. A larger overlap leads to quantitative changes and the loss of a
mirror symmetry of pairs of energy eigenvalues in the $(\mu, \gamma)$-diagram,
however, it does not affect the generic structure of the states.

The influence of the phase difference between the two subsystems was also
examined.  While the coupling strength $\gamma$ between the two subsystems took
the role of one bifurcation parameter, the phase difference $\varphi_{\rm rel}$
took the role of another, leading to a two-parametric cusp bifurcation. This
bifurcation degenerated for $\varphi_{\rm rel} = 0$ into a pitchfork
bifurcation. Only in this case \PT-symmetric states could be observed which
makes the phase difference between the subsystems critical for the
\PT-symmetric properties of the system.

The matrix model can be investigated further. Under the assumption that the two
wells of the system are sufficiently isolated the matrix model reduces the
description of the system to a low number of key parameters.  Therefore the
analytically accessible matrix model of this paper could be helpful to gain
more insight into the behaviour of coupled BECs. In particular a similar
approach to realize a \PT-symmetric quantum system via the coupling of two
condensate wave functions was studied in \cite{Single2014} and revealed
complicated stability properties. This system should also be representable in
our four-mode description such that analytic expressions should be obtainable.

\ack

This work was support by Deutsche Forschungsgemeinschaft.

\appendix

\section{Time-independent solutions of the nonlinear GPE}
\label{s:tigpe}

We will first consider a linear GPE in appropriate units
\begin{equation}
	\ii \frac{\partial}{\partial t} \psi = - \Delta \psi + V(x) \psi.
\end{equation}
To find time-independent solutions one uses
\begin{equation}
	\psi(t) = \psi_0 \exp(-\ii \mu t) \label{eq:tiansatz}
\end{equation}
which leads to the time-indepedent equation
\begin{equation}
	\mu \psi_0 = -\Delta \psi_0 + V(x) \psi_0.
\end{equation}
Solutions with a real $\mu$ are true stationary states, i.e. only the global
phase is changed with $\exp(-\ii\, \Re \mu\, t)$. By contrast states with a
complex $\mu$, in addition to the phase change, increase or decrease
in the probability amplitude exponentially with $\exp(\Im \mu\,t)$.

For a GPE with a nonlinearity this is no longer true. If we consider 
\begin{equation}
	\ii \frac{\partial}{\partial t} \psi = - \Delta \psi + V(x) \psi +g  | \psi |^2 \psi
\end{equation}
the previous ansatz \eref{eq:tiansatz} will lead to
\begin{equation}
	\mu \psi_0 = -\Delta \psi_0 + V(x) \psi_0 + g | \psi_0 | \psi_0 \exp(-2 \ii \,\Im
	\mu \,t).
\end{equation}
Also in this case, if $\mu$ is a purely real number, the states $\psi$ are
stationary. But for states with a chemical potential $\mu$ which has an
imaginary part $\Im \mu \neq 0$ the interpretation changes. In the nonlinear
case \eref{eq:tiansatz} is only a solution in the limit $t \rightarrow 0$.
Therefore for small times the probability amplitude of these states still
approximatially increases or decreases exponentially, but the true time
evolution deviates from this linear solution as time increases.

\section{Analytical solutions of the matrix model}
\label{s:analyticalsolutions}

We want to show how to calculate the analytical solution for the four
dimensional matrix model in \eref{eq:complexmu}. As an ansatz for \PT-symmetric
solutions \eref{eq:ptsymansatz} is used.  We obtain the equations
\begin{eqnarray}
	-\frac{g}{2} + v\ee^{-2\ii\varphi} - \ii\gamma\ee^{-2\ii\varphi} &= \mu, \nonumber \\
	-\frac{g}{2} + v\ee^{2\ii\varphi}  + \ii\gamma\ee^{2\ii\varphi}  &= \mu.
	\label{eq:ptsymeq}
\end{eqnarray}
With the substitution $x = \exp(2\ii\varphi)$ we can transform these equations
into
\begin{eqnarray}
	(v - \ii\gamma) \frac{1}{x} = \mu + \frac{g}{2}, \nonumber \\
	(v + \ii\gamma) x = \mu + \frac{g}{2}.
\end{eqnarray}
This in turn leads to
\begin{eqnarray}
	(v - \ii\gamma) \frac{1}{x} = (v + \ii\gamma) x
\end{eqnarray}
and therefore we obtain
\begin{eqnarray}
	x = \pm \sqrt{\frac{v-\ii\gamma}{v+\ii\gamma}},
\end{eqnarray}
which can be inserted into one of the equations in \eref{eq:ptsymeq} and yields
to the two \PT-symmetric solutions
\begin{eqnarray}
	\mu = - \frac{g}{2} \pm \sqrt{v^2 + \gamma^2}.
\end{eqnarray}

For the \PT-broken solutions the ansatz \eref{eq:ptbrokenansatz} is used and
results in
\begin{eqnarray}
  -g \cos^2\theta + v\tan\theta\ee^{-2\ii\varphi} -\ii\gamma\ee^{-2\ii\varphi}
			&= \mu, \nonumber \\
	-g \sin^2\theta	+ v\cot\theta\ee^{2\ii\varphi} + \ii\gamma \ee^{2\ii\varphi}
			&= \mu.
			\label{eq:ptbrokeq}
\end{eqnarray}
By eliminating $\mu$ and separating the equation into its real and imaginary
part the equation system
\begin{eqnarray}
	-g (\cos^2\theta-\sin^2\theta)+v(\tan\theta-\cot\theta)\cos2\varphi &= 0 \nonumber \\
	-v(\tan\theta + \cot\theta)\sin2\varphi - 2\gamma\cos2\varphi &= 0
\end{eqnarray}
remains, which can be transformed into
\begin{eqnarray}
	\sin2\theta &= -\frac{2v}{g} \cos2\varphi = -\frac{v}{\gamma} \tan2\varphi.
\end{eqnarray}
By the substitution $x = \exp(2\ii\varphi)$ the quasi palindromic polynomial
\begin{eqnarray}
	x^4 - 2A x^3 + 2 x^2 + 2A x +1 = 0 \quad{\rm with}~ A = -\ii\frac{g}{2\gamma}
\end{eqnarray}
is obtained. The four solutions of this polynomial are
\begin{eqnarray}
	\fl
	x = \frac{1}{2}(z\pmsb\sqrt{4+z^2}) \quad{\rm with}~
	z = A \pmsa \sqrt{A^2 -4} = -\frac{g\ii}{\gamma} P \quad{\rm with}~
	P = \frac{1}{2} \pmsa \frac{\sqrt{g^2 + 16 \gamma^2}}{2g}.
\end{eqnarray}
Note that the $\pm$ for $x$ and $P$ are independent and therefore lead to four
solutions.  By inserting the solutions into one of the equations in
\eref{eq:ptbrokeq} one obtains the analytical expressions for the chemical
potential,
\begin{eqnarray}
	\mu = -\frac{g}{2} \left( 2 \mpsb \sqrt{P+\frac{\gamma^2}{v^2}P^2}-P \right).
\end{eqnarray}
Note that without an analytical continuation of the equations the parameters
$\theta$ and $\varphi$ in the ansatz of the wave functions must be real.
Therefore one can see that two of the solutions for the chemical potential have
complex $\theta$ or $\varphi$ over the whole parameter range and therefore are
shown in lighter colours in \fref{fig:matrix}. The other two solutions exist if
the constraint
\begin{eqnarray}
	\gamma > \gamma_c = \sqrt{\frac{4v^4}{g^2}-v^4}
\end{eqnarray}
is fulfilled.

For the effective matrix model in \eref{eq:ef2dmatrix} with the ansatz
\begin{eqnarray}
	\psi = \left( \cos\theta, \sin\theta \right)
\end{eqnarray}
one obtains the equations 
\begin{eqnarray}
	-g \cos^2\theta - \gamma + v\tan\theta &= \mu, \nonumber \\
	-g \sin^2\theta + \gamma + v\cot\theta &= \mu,
\end{eqnarray}
which can be transformed into the polynomial
\begin{eqnarray}
	g y^4 + 4(\gamma + \ii v) y^3 + 4(-\gamma +\ii v)y -g = 0
\end{eqnarray}
by eliminating $\mu$ and substituting $y = \ee^{2\ii\theta}$. The four
solutions of the polynomial can be obtained by any of the methods to solve
polynomials of degree four. Once they are known $\mu$ can be calculated.

\section{Derivation of the coefficients in the matrix model from the extended
Gaussian model}
\label{s:matrixmodelderivation}

The matrix model \eref{eq:matrix4} can be derived as a discrete nonlinear ansatz
of the extended model (for the derivation of a nonlinear discrete Schr\"odinger
equation from the GPE see \cite{Tromb2001,Smerzi2003,Kreibich2014}). We
rearrange the terms in \eref{eq:gauss_tdvp} which results in
\begin{eqnarray}
	i \dot\psi_{\rm A} &=
		\Big(
			\underbrace{-g|\psi_{\rm A}|^2}_{H_{A0}}
			\underbrace{-\partial_x^2+\frac{1}{4} x^2 + V_0^{\rm G} \ee^{-\sigma x^2}}_{H_{1}}
		\Big) \psi_{\rm A}
		\underbrace{+\ii\gamma x \ee^{-\rho x^2}}_{H_{\rm AB}} \psi_{\rm B},
	\\
	i \dot\psi_{\rm B} &=
		\Big(
			\underbrace{-g|\psi_{\rm B}|^2}_{H_{B0}}
			\underbrace{-\partial_x^2+\frac{1}{4} x^2 + V_0^{\rm G} \ee^{-\sigma x^2}}_{H_{1}}
		\Big) \psi_{\rm B}
		\underbrace{-\ii\gamma x \ee^{-\rho x^2}}_{H_{\rm BA}} \psi_{\rm A}.
		\label{eq:new_gauss}
\end{eqnarray}
Also a slightly different parametrisation for the ansatz of coupled
Gaussians \eref{eq:gauss_ansatz} is used
\begin{equation}
	\psi_{\rm i}
	 = \sum_{j=1,2} g_{i,j}
	 = \sum_{j=1,2} d_{i,j}(t) f_{i,j}(x)
	 = \sum_{j=1,2} d_{i,j}(t) \ee^{a_{i,j}(x-q_{i,j})^2 + p_{i,j}(x-q_{i,j})} \label{eq:new_ansatz}
\end{equation}
with $i = A, B$, $j=1,2$, $a_{i,j} \in \mathbb{C}$ and $p_{i,j},q_{i,j} \in
\mathbb{R}$. In this new ansatz the amplitude and phase $d_{i,j}$ is separated
from the shape $f_{i,j}$ of the wave functions. It is assumed that the shape is
constant in time and only the amplitude and phase changes.

In the following paragraphs we only consider the equation of subsystem A the
calculation for subsystem B can be done in the same way. We insert the new
ansatz \eref{eq:new_ansatz} into \eref{eq:new_gauss} and obtain
\begin{equation}
	\sum_{k=1,2} \ii \dot d_{A,k}
		= \sum_{k=1,2} (H_{\rm A} + H_1) d_{A,k} f_{A,k} + H_{\rm AB} d_{B,k} f_{B,k}.
\end{equation}
The equation is multiplied with $f_{\rm A,1}^*$ and $f_{\rm A,2}^*$ from the
left and the equation is integrated over $x$.
The resulting two equations can be written as a matrix equation
\begin{eqnarray}
	\fl	\ii \underbrace{\left(\begin{array}{cc}
		\langle f_{A,1} | f_{A,1} \rangle & \langle f_{A,1} | f_{A,2} \rangle \\
		\langle f_{A,2} | f_{A,1} \rangle & \langle f_{A,2} | f_{A,2} \rangle
	\end{array}\right)}_{=K_{\rm A}} \underbrace{\left(\begin{array}{c}
		\dot d_{A,1} \\ \dot d_{A,2}
	\end{array}\right)}_{=\dot d_{\rm A}} \nonumber \\
	=
	\underbrace{\left(\begin{array}{cc}
		\langle f_{A,1} |H_{\rm A}+H_1| f_{A,1}\rangle & \langle f_{A,1} |H_{\rm A}+H_1| f_{A,2} \rangle \\
		\langle f_{A,2} |H_{\rm A}+H_1| f_{A,1}\rangle & \langle f_{A,2} |H_{\rm A}+H_1| f_{A,2} \rangle
	\end{array}\right)}_{=G_{\rm A}} \left(\begin{array}{c}
		d_{A,1} \\ d_{A,2}
	\end{array}\right) \nonumber \\ + \underbrace{\left(\begin{array}{cc}
		\langle f_{A,1} |H_{\rm AB}| f_{B,1}\rangle & \langle f_{A,1} |H_{\rm AB}| f_{B,2} \rangle \\
		\langle f_{A,2} |H_{\rm AB}| f_{B,1}\rangle & \langle f_{A,2} |H_{\rm AB}| f_{B,2} \rangle
	\end{array}\right)}_{=G_{\rm AB}} \left(\begin{array}{c}
		d_{B,1} \\ d_{B,2}
	\end{array}\right).
\end{eqnarray}
Combining the equations from both subsystems results in a four-dimensional
matrix equation
\begin{equation}
	\ii \left(\begin{array}{cc}
		K_{\rm A} & 0 \\
		0 & K_{\rm B} 
	\end{array}\right)
	\left(\begin{array}{c}
		\dot d_{\rm A} \\ \dot d_{\rm B}
	\end{array}\right)
	=
	\left(\begin{array}{cc}
		G_{\rm A}  & G_{\rm AB} \\
		G_{\rm BA} & G_{\rm B} 
	\end{array}\right)
	\left(\begin{array}{c}
		d_{\rm A} \\ d_{\rm B}
	\end{array}\right).
\end{equation}

We add a numerical example, which is obtained for $g = 0.2$ and $\gamma=0.03$
for the extended model (compare \fref{fig:gaussmatrix}).

First we examine the matrices $K_{\rm A}$ and $K_{\rm B}$. 
\begin{equation}
	K_{\rm A} = K_{\rm B} = \left(\begin{array}{cc}
		1.87 & 0.027 - 0.0097 \ii \\
		0.027 + 0.0097 \ii & 1.87
	\end{array}\right).
\end{equation}
It is obvious that the matrix has only small off-diagonal elements since the
overlap of the wave functions of different wells is very small. Therefore the
matrix $K$ can be approximated by a diagonal matrix $D$. Then the equation is
multiplied with $D^{-1}$ from the left.

Now we examine the matrix elements of the matrix $G_{\rm AB}$. The
first diagonal element of the matrix is
\begin{equation}
	-\ii\gamma \langle f_{\rm A1}^* | x \ee^{-\rho x^2} | f_{\rm B1} \rangle,
\end{equation}
where the second term in brackets contains only structural information and
can be integrated. It corresponds to the fit parameter $\gamma_0$. We examine
the numerical values of the matrix
\begin{equation}
	G_{\rm AB} = 
	\left( \begin{array}{cc}
		9.59\times10^{-4} + 6.31\times10^{-2}\ii & 9.40\times10^{-5} \\
		9.40\times10^{-5} & 9.59\times10^{-4} + 6.31\times10^{-2}\ii
	\end{array} \right).
\end{equation}
It is obvious that the small overlap of the wave functions in different wells
leads to very small off-diagonal elements, which can be neglected.

The entries of the matrices $G_{\rm A}$ and $G_{\rm B}$ consist of terms
containing the external potential and the kinetic energy on the one hand
and terms containing the contact interaction on the other hand.
\begin{table}
\caption{\label{tab:matrixnum} Numerical absolute values of the matrix entries
	for matrices $G_{\rm A}$ and $G_{\rm B}$ for $g =0.2$ and $\gamma = 0.03$. See
	\fref{fig:gaussmatrix}.}
\begin{tabular}{c||c|c}
	 & contact interaction  & external potential and kinetic energy\\
	 & $H_{\rm A0}$ and $H_{\rm B0}$ & $H_1$ \\
	\hline
	diagonal &
	$7.1214 \times 10^{-2}$ &
	$4.6221$
	\\
	off-diagonal &
	$2.3891 \times 10^{-4}$ &
	$8.2474 \times 10^{-2}$
\end{tabular}
\end{table}
The terms of the external potential and the kinetic energy in the diagonal
element induce a shift of the energy (which corresponds to the offset $\Delta
\mu$ in the fit), therefore only the nonlinear contact interaction term remains
on the diagonal. The contact interaction in the off-diagonal is very
small (compare \tref{tab:matrixnum}) when compared to the diagonal, and can be
neglected. Therefore in the off-diagonal only the terms from the external
potential and the kinetic energy remain. These correspond
to the coupling parameter $v$ in the fit.

\section*{References}
\bibliographystyle{unsrt}
\bibliography{paper}

\end{document}